# BUDAMAF

*Data Management in Cloud Federations*

Evangelos Psomakelis[1,2], Konstantinos Tserpes[1,2], Dimosthenis Anagnostopoulos[1] and Theodora Varvarigou[2]

[1]*Dept. of Informatics and Telematics, Harokopio University of Athens, Omirou 9, Tavros, Greece*
[2]*Dept. of Electrical and Computer Engineering, National Technical University of Athens, Iroon Polytecniou 9, Zografou, Greece*
vpsomak@mail.ntua.gr, tserpes@hua.gr, dimosthe@hua.gr, dora@telecom.ntua.gr

Abstract: Data management has always been a multi-domain problem even in the simplest cases. It involves, quality of service, security, resource management, cost management, incident identification, disaster avoidance and/or recovery, as well as many other concerns. In our case, this situation gets ever more complicated because of the divergent nature of a cloud federation like BASMATI. In this federation, the BASMATI Unified Data Management Framework (BUDaMaF), tries to create an automated uniform way of managing all the data transactions, as well as the data stores themselves, in a polyglot multi-cloud, consisting of a plethora of different machines and data store systems.

## 1. INTRODUCTION

### 1.1 BASMATI

BASMATI (CAS Software AG, 2016) is a cooperative project between Europe and Korea, aiming at the creation of a cloud federation platform that can easily host cloud applications. It will provide the ability of scaling between multiple cloud providers, based on the requested QoS as well as the pricelist of each provider supported. This creates not only a vast, federated pool of resources but also an automated process of finding the most cost effective solution by combining resources from multiple cloud providers (CPs).

For example, a user in Korea could be using a virtual desktop service, where a virtual machine he is managing is hosted in the cloud. He needs to be in close proximity to the cloud datacenter because in this application gigabytes of data are flowing between the datacenter and the user's laptop. If this user is now traveling to Europe for business, he will encounter a huge delay when he tries to access his virtual desktop again because of all the distance between him and the Korea hosted datacenter. With BASMATI, the user's data would travel at the same time as he to a European hosted datacenter, controlled by another CP that is part of the federation, even if the application provider (AP) has no idea which CP is that. So not only will he always be in close proximity to his data, but the cost to the AP will have small variance between the European and Korean datacenters, following automated procedures of cost analysis.

In order to achieve that, BASMATI uses an automated SLA negotiation process that predefines all the costs for resource allocation as well as the federated resources that each CP provides. Then, using the ACE (Marshall, 2016) system, it creates a common resource pool, allowing a CP that is unable to serve an application deployment request (either due to low resource availability or to QoS requirements such as the user location or the need of specific technologies), to automatically "borrow" resources, in real time, from another CP that is part of the federation.

### 1.2 BASMATI Data Management Problems

#### 1.2.1 Polyglot Persistence

BASMATI aims to provide a context irrelevant solution, supporting any application that can be deployed on a cloud. In order to achieve that, it needs to create methodologies that support all platforms and data stores. Since we are dealing with applications already developed, we can expect each application to use its own data store system, according to its needs and the personal choices of its developers.

This creates a problem known as Polyglot Persistence (which will be defined in a later section in detail). In short, we cannot manage any data store system by using the same data management system. For example, an SQL system follows a certain syntax, providing specific functionality. We cannot expect a NO-SQL system, like MongoDB, to provide the same functionality or to use the same syntax. On the other hand, an application using BASMATI should not care what database it is using, it should make uniform data requests, no matter what data store they are targeting.

Another reason making the polyglot persistence a necessity is having jobs that need to have a holistic view of the data. BUDaMaF will provide support for data analytics jobs, using the data stored in its databases, data that in some cases come from the applications themselves and thus, they originate from different data store systems. The analytics jobs will be using data from all these varied sources, so they will need a uniform way of accessing these sources.

### 1.2.2 Data security and privacy protection

Another big problem is the data security and protection of privacy. In a cloud federation, data are always travelling all around the world, from CP to CP in order to ensure maximum performance and lowest costs. This creates many opportunities for third parties to access or corrupt these data along the way, causing problems to the security or the basic functions of hosted applications.

### 1.2.3 Resource monitoring and bottleneck avoidance

When managing data in a complex system, like a multi-cloud federation, we have to keep a close eye on resource availability. Each CP provides us with a number of available resources, according to the SLAs signed. This can create bottlenecks when the demand is increasing suddenly in certain cases. A smart mechanism needs to be in place to avoid or handle this problem in real time, by off-loading the increased system load or scaling the applications in the federation.

### 1.2.4 ACID enforcement

As discussed earlier, each data store in the federation may or may not support ACID properties. Regardless of that, the framework as a whole needs to provide it in order to cover the needs of the applications that need these properties in their transactions. Even if a hosted application does not need ACID properties, it will not "hurt" its functionality, in an average case, if the system provides them either way. Of course, in order to provide these properties even in NO-SQL data stores, a controller responsible for this needs to be created, forming a level of abstraction between the query engine and the data stores actually used.

## 1.3 BUDaMaF Goals

As a part of a larger system, BUDaMaF is trying to cover the needs of BASMATI. Because of that it strives to cover basic data management tasks (read, write, update, delete) and basic data store tasks (creation, automated horizontal scaling, deletion, relocation) before everything else. Once these are secured, the secondary goals of BUDaMaF include federation specific data requests (mitigation, off-loading, replication), security and privacy insurance tasks (protocol enforcement, access control, encryption/anonymization enforcement) and ACID enforcement.

### 1.3.1 Basic Goals

The goal for the basic data management tasks, as mentioned earlier, is to simply create a query processor that can understand what the user needs by reading a JSON format query and forward this query to the responsible component of the framework, as we will see later in the architecture of BUDaMaF. For the data stores now, a more sophisticated middleware needs to be created in order to serve as a dashboard that a user (be it a person or an application) can use to perform the tasks mentioned earlier on any data store that is supported by the framework, in an automated (or semi-automated if that is not possible) way.

### 1.3.2 Federation Data Requests

As discussed, three advanced request types can be served inside the framework; mitigation, replication and off-loading. Data mitigation is needed in cases that data need to move from one data store to another or from one physical data center site to another. This commonly is becoming a necessity due to cost limitations or datacenter proximity to the user or to the application provider.

Data replication is about copying the data from one data store to another or from one physical data center site to another. The simplest cause for that request is the need to create a backup. In other cases, the user may need the data to be available and synchronized in two different data centers or data stores. For example if two people on different parts of the world (one in Korea and one in Europe for example) are working together on a project and need access to the same data, without having to endure vast delays in response times.

Finally, off-loading is about the case that an application is creating huge amounts of load at specific times on the data store. At these times, the data requests need to be off-loaded, either by redirecting the requests to other data stores able to serve them or requesting real-time data store scaling.

#### 1.3.3 Security and Privacy

The framework is designed having in mind that an external source will provide the guidelines for security and privacy management of the data. This is derived from the assumption that security protocols and privacy protection guidelines are always evolving, so a detached security and privacy authority needs to keep them always updated.

Having that in mind, BUDaMaF has a component that acts both as a dashboard for a security administrator that can define security protocols and privacy guidelines and as a centralized authority that the framework can use to coordinate the security and privacy enforcement actions. Each time a new data request is directed to the framework Core, a request is made to the security and anonymization engine in order to identify what level of security, access control and privacy protection is needed for that request and what protocols need to be followed.

#### 1.3.4 ACID enforcement

ACID enforcement is not currently part of the framework, even though it is marked as a future goal, amongst others that will be presented in a following section. The way requests are currently handled in BUDaMaF, ACID is already ensured for data stores that were providing it by default, so if an application was already using a data store with ACID support, that support will remain in the framework as long as the same data store types are being used for this application.

That is the case because the framework is serving data requests by forwarding them to a responsible wrapper which, in essence, is just an automated client to the target data store. For example, if an SQL query is made to the framework, it will search for an attached SQL Wrapper and forward this query. The Wrapper will execute the query in the target SQL data store and receive the response. Then it will forward the response to the Core and it will present it to the request initiator. More details about that process will be provided in a following section.

### 1.4 Impact

This paper presents the architecture, interfaces, inner workings and the use cases of a novel multi-cloud federation data management framework called BUDaMaF. This framework was developed in the context of a cooperative Korean – European project called BASMATI in order to cover its data management needs, but as we will discuss in this paper, it is not limited to BASMATI.

BUDaMaF is a generalized framework, able to handle the needs of many platforms for cloud federation management. It can even be used as a separate entity, providing a federation data management dashboard, covering all the needs, from data store management to high level data management, while providing polyglot persistence.

As a context agnostic framework, it is not bound by specific domains, platforms or even underlying technologies. All it needs to be compatible with any system is the appropriate Wrappers, that can be developed using any technology, by any person and then attached to the framework using the loosely coupled architecture of RestFul Web Services.

Finally, due to the clear separation of its basic functionalities into singular components, BUDaMaF is highly scalable. It can be deployed either in a single machine, in a cluster of machines or in a cloud with no additional effort, making it an ideal solution for cloud federations with irregular data traffic loads.

## 2. RELATED WORK

### 2.1 Cloud Federations

The world of cloud computing is full of market shares and competitive corporations. In order to provide a cloud service, a CP needs to invest huge amount of money in data centers and computing resources, regardless of the type of cloud services provided; IaaS, SaaS or PaaS. Today, the market is conquered by four major providers, Amazon (AWS), Microsoft (Azure), Google (Google Cloud Platform) and IBM (Bluemix).

In Figure 1 we can see the market shares of the dominant CPs. As we can see Amazon is the greatest provider with Microsoft, Google and IBM following in second place. But all this competition to conquer the market and get some profit from the investment made to buy all the resources makes us forget about the basic principle of cloud computing, which is pooling of resources.

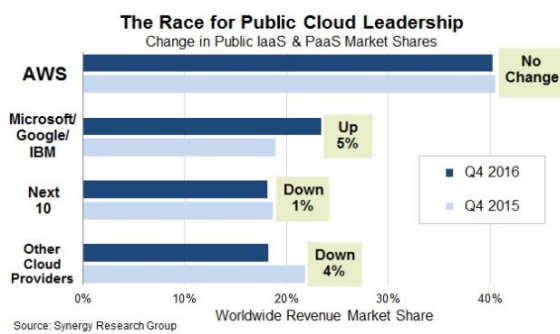

Celesti et. al. remind us of this fact by clarifying Figure 1: Market Shares for Cloud Providers (Panettieri, 2017).
the three stages of cloud federations (Celesti et al., 2010). They mention that currently most CPs are built based on stage 1 of cloud computing which is the most primitive stage of relying only on resources owned by one provider. Stage 2 is an evolution of that where each provider still holds tight to its resources but also buys resources from other providers if it suits its needs. Stage 3 on the other hand is creating a common pool of resources by regarding all resources, both the ones owned by the CP and the ones rented from other CPs, as the same.

Rochwerger et. al. are concerned with another aspect of cloud federations, the limited interoperability that CPs are providing (Rochwerger et al., 2011). They are also trying to implement a system that creates a common pool of resources between several CPs. They mention several concerns that need to be addressed in such an effort, revolving around the optimization of resource usage and cost efficiency. Above all else, they mention that the CPs have created their systems without thinking about interoperability, which makes a middleware necessary in order to provide a level of abstraction to the deployment process.

On the other hand, a new solution may be emerging to tackle the interoperability issues of cloud federations. Open Cloud Computing Interface (OCCI) is an open standard as well as a working group improving three basic aspects of cloud IaaS services; portability, interoperability and integration (Metsch, 2006). It aims to achieve this by providing a slim (about 15 commands) RestFul API for IaaS management, including resource and virtual machine management, based on the HTTP and other, already established, standards.

Since its creation, OCCI has gathered a lot of support by academia and already has implementations for an impressive number of cloud management systems such as OpenStack, CloudStack, OpenNebula, jClouds, Eukalyptus, BigGrid, Okeanos, Morfeo Claudia and others (OCCI-WG, n.d.).

## 2.2 Cloud Data Management

### 2.2.1 Basic Principles

Tuan Viet lists eight basic principles of cloud computing that are mentioned for a more complete view of the cloud technology (Viet-DINH, 2010);
- Resource Sharing: One of main attributes of clouds is their ability to provide on demand resources to a user, when she requires them.
- Heterogeneity: A computer cloud is consisted of various physical machines, using different technologies, both in software and hardware, creating a heterogeneous system.
- Virtualization: All this heterogeneity would make the usage of this system very complicated even for trained users. Virtualization masks all that by rearranging the resources and providing a virtual environment that the user can manage and use with more ease.
- Security: The remote nature of the resources used give rise to a number of security threats, such as man in the middle and denial of service attacks. For that reason, security is a main concern in any cloud application.
- Scalability and Self-Management: Another main feature of clouds is providing the ability to increase or reduce the resources of their virtual machines after their original creation, in other words to scale their machines. This operation requires the cloud to self-manage its physical resources in order to reallocate them, as needed, in an unsupervised way.
- Usability: Clouds need to be usable both by experienced and by casual users, which requires them to hide a lot of their inner workings and configuration options under layers of background processes.
- Payment Model: Cloud providers need to cover their investment in infrastructures, both for the initial purchase and for the maintenance and upkeep costs. That is the reason that most cloud providers use specific and pre-agreed payment models that connect either the time of usage or the resources available to the users with monetary costs.
- Quality of Service: Clouds are above all software products and as such a certain QoS threshold need to be established in an SLA contract between the cloud provider and the user.

### 2.2.2 Major Cloud Provider Solutions

As discussed in a previous section, Amazon is the most popular cloud provider in the market. They are providing two services for data management using their cloud infrastructure, one for relational databases and one for non-relational ones. In this paper we will focus on the non-relational one which is closer to our big data needs. The service mentioned is called Amazon Elastic MapReduce (EMR) (AWS, n.d.). It is using the cloud infrastructure of Amazon to provide big data management solutions in many of the most popular data store systems of the market, including HDFS, Presto, Spark and others.

Google, another major player in the cloud computing market, has created their custom solution for data management called Dremel (Melnik et al., 2010). The relevant software provided to the users though is called BigQuery, which is actually based on the Dremel software (Sato, 2012). Sato mentions that Dremel is complementing the classical MapReduce by improving the seek time, making it possible to execute a read query in a 35.7 GB dataset in under 10 seconds. Also, all this is done using classical SQL queries, so Google actually created a powerful, scalable data management solution without the need to create a new query language.

The third great provider is Microsoft, with their Azure Cosmos DB software (Shukla, 2017). Cosmos DB is a cloud data base engine based on atom-record-sequence (ARS). This enables it to function as an extremely scalable data management engine while providing support to multiple popular data management systems like DocumentDB SQL, MongoDB and Gremlin. It also provides easy API access to most programming languages, using simple JSON representations, enabling users to access its functionality from their customized clients.

### 2.2.3 Polyglot Persistence

When talking about traditional data management, we imagine a data administrator managing a database in a high end data server. This image gets more and more obsolete as the cloud technology and parallel computing are advancing, both in technological level and in low cost solutions. Regardless of all the advances of such a solution, including cost effectiveness and easier scalability, a new problem arises, that of the polyglot data stores.

A cloud consists of many machines and many different data store systems, either due to machine limitations or due to the need for specialized tasks (Kolev et al., 2015). When trying to use these different data store systems in an interconnected cloud we encounter the polyglot persistence (Fowler, 2011) problem, which tries to manage a group of different data stores, talking in different languages, by using a common interface.

A common solution to that problem is creating wrapper components that translate queries from a common language to the native language of each different data store (Bondiombouy, 2015; Bondiombouy et al., 2015; Kolev et al., 2015, 2016; Zhu and Risch, 2011). They also need to translate the response into a format readable by the common language processor. That methodology permits both the support of new data stores as needed, just by creating the corresponding wrapper, and the definition of new query types in the common language.

The major drawback in this methodology is that this new language is not standardized, so any researcher develops his/her own language that is incompatible with all the others (Wang et al., 2017). That means extra hours of training for anyone who wants to use this polyglot persistent system. The solution commonly followed, in order to counter this problem, is using an SQL based language and adding specialized commands on it by either expanding its glossary or using PLSQL (Bondiombouy et al., 2015; Kolev et al., 2016; Zhu and Risch, 2011).

## 3. ARCHITECTURE

## 3.1 General Architecture

BUDaMaF is comprised of four basic components, as seen in Figure 2. These are the Core, the Analytics Engine, the Off-Loading APIs and the Security and Anonymization Engine. Each component will be presented in detail in the following sections. For now let's consider them "black boxes" and comment on the general architecture of the framework.

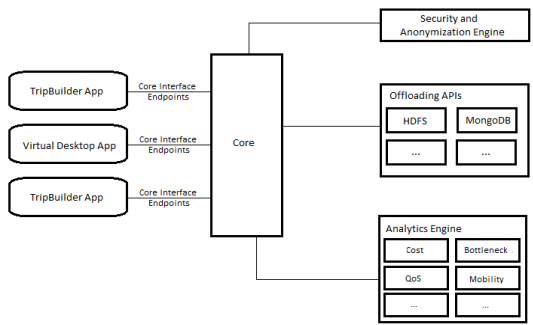

Figure 2: BUDaMaF General Architecture.

The general architecture diagram, presented in Figure 2, shows that each one of the four main components is responsible for a single role that the framework needs to play, with the exception of the Core components, which acts as the coordinator between all the components. The Off-loading APIs component handles the data store and data management requests, the Analytics Engine handles the data mining and data analytics requests and the Security and Anonymization Engine handles all the security concerns that arise due to the vulnerable nature of the internet communications between the federation members and their resources.

All the requests are passing through the Core component, which then decides how to handle each request. This component then redirects the request to the responsible component, lets it handle it and then receives the response. The response then is forwarded to the original request owner, except if the response demands an action from the Core first. For example, if the response reads as a timeout error, the Core may redirect the original request to another instance of the responsible component, in order to overcome this error before responding to the request owner.

## 3.2 Components

### 3.2.1 Core

The Core component acts both as the coordinator for the other components and as a portal for the outside world. All the requests and data are passing through the Core before being pushed in or out of the framework. This enables the Core to communicate with the Security and Anonymization Engine in order to get information about which protocols to enforce on the data and actually enforce them on all data passing through the BUDaMaF. It also enables the users (be it actual people or pieces of software) to access the framework in a uniform way, using one access point (which can be scalable in order to avoid bottlenecks) and a common request glossary in JSON format, regardless of the job they need to perform.

The Core can also keep information about requests, giving us the ability to enforce access control to the data, enhancing data security. For example, the data created by one of the applications hosted in BASMATI would be accessible by the application that created them and a limited set of analytics components that the application chooses to grant access rights. When the Core receives a data request it can decide if the request owner has access rights to this kind of data and then take the necessary actions to either provide the data or inform the request owner that access to these data is prohibited.

The final job of the Core component is load balancing and delay handling. It can identify bottlenecks and try to avoid them by using alternative component instances if it can find any. If no instances are available it can inform the request owner that the framework suffers from a high load of requests and patience is advised while logging the issue and notifying the responsible administrators.

### 3.2.2 Security and Anonymization Engine

This component supports the BUDaMaF by providing security and anonymization guidelines concerning the data flowing through the framework. It takes into account the current technologies, as described in its option which are set by a security administrator, the owner of the data, as described in the metadata travelling with the data, and the type of the data, as described in the metadata as well.

In the framework we have distinguished three kinds of data: the federation data, the application data and the monitoring data. The federation data are data created and used by BASMATI itself, aiding its processes. It can include request metadata, resource information, cost functions and many other information. These data are not deemed as highly private so no special anonymization or security is needed, with the exception of integrity preservation. The monitoring data include already anonymized monitoring data, gathered by the hosted applications or the federation machines. They are used mainly for analytics and optimization tasks. Given that they are already anonymized, again no special security measures need to be taken except of integrity preservation.

On the other hand, the application data, which is the third category, contains private data such as user names, age groups, trajectories, favorite places, photos, personal documents and many others. As such, we need to safeguard the privacy of these data by anonymization wherever possible and encryption when needed. This will ensure that even if the basic access control implemented by Core fails and the data end up in the wrong hands, they will be either unreadable or anonymized.

### 3.2.3 Off-Loading APIs

As discussed, one of the main targets of this framework is to manage not only the data traversing through the federation but also the data stores themselves. In order to do that in a unified way, we need a connector that translates the common requests into more specific commands for each one of the different data stores hosted in BASMATI.

For that purpose we have the Off-Loading APIs component, which receives data and data store requests in a uniform way, following the OCCI specifications. Then it decides who is responsible for handling this request and forwards it to the responsible Wrapper module. This process ensures that regardless of the data store system and the technologies involved in the machines that host this data store, the interface and the commands to manage this data store or the data contained in it are constant.

The Wrapper modules are smaller components that are developed separately and they are loosely coupled to the Off-Loading APIs component through RestFul Web Services. This architecture enables individual developers or even users of the framework to develop their own module, adding support for their preferred data store. This module is responsible for translating the high level requests received from the Off-Loading APIs component into commands that make sense to the target data store and then translating the response into a format acceptable by the Off-Loading APIs.

The requests this component handles are basic data requests (read, write, update, delete), federation specific data requests (mitigation, replication, publication, anonymization, encryption) and data store requests (horizontal scaling, relocation, creation, destruction). More details on these requests will be provided in section 3.3 Interfaces.

### 3.2.4 Analytics Engine

This component will handle all data requests by analytics modules. It will not perform analytics tasks itself. Instead, it will act as an access point for the specialized analytics modules, loosely coupled to it, using RestFul Web Services. Each module will be responsible for one specific analytics task in a specific dataset, which may contain data from multiple sources, be it internal or external to the federation.

This enables the Analytics Modules to find the data they need without caring about where or how they are stored in the federation. They can just locate and access them by describing them, using a common interface and a common glossary in JSON format.

The analytics modules can be developed separately and then connected to the analytics engine by using its access point. This way each new analytics module will enhance the range of tasks the framework can perform, giving it the ability to serve even more requests as it evolves.

## 3.3 Interfaces

Each component of the framework exposes several RestFul APIs, following the OCCI standard. These interfaces are built in a way that each component is an OCCI category, having specific actions and attributes, while using the standard http methods, as the OCCI standard dictates. In the rest of the chapter we will present the various interfaces of the framework.

### 3.3.1 Common Specifications

As all components are parts of the same framework, some specifications are common but not shared. Each component is using the specifications for its own needs. Though, in order to save space in this paper, we decided to present all the common specifications in this section instead of presenting the same things over and over again. If any deviation is present in a specific component interface it will be mentioned in that component's sub-section.

**Members**:
- initiator: the initiator credentials for this job in JSON format.
- job_description: The type of this job.
- job_details: The details for this job in JSON format.
- status: The status of this job, either pending, running, finished or crashed.
- data: The provided data in insertion jobs or a placeholder for the requested data in retrieval jobs or an error description if an error is encountered.

**Methods**: The standard OCCI methods are managed for this category endpoint with the following specificities.

GET

This method may be used to retrieve the data requested if the instance was started with a retrieval job or the status of a request if no data are to be retrieved.

PUT

This method is used in job requests that need to keep an open channel, accepting data while they are active, for the other job types all the data are included in the original POST request that creates them.

DELETE

This method may be used to delete the instance and release resources.

### 3.3.2 Core

**Category Name**: core
**Description**: Instances of this category will be created in order to start generalized jobs in the BASMATI Unified Data Management Framework (BUDaMaF), interconnecting and managing the individual sub-components.
**Links**: Core instances will be linked to the Offloading APIs and the Hosted Applications.
**Methods**: The standard OCCI methods are managed for this category endpoint with the following specificities.

POST

This method will be used to initiate a job in the BUDaMaF by providing the initiator id, the type and the details of the job. The type can be the name of any supported operation, whereas the details depend on the job type, some jobs need only a minimal quantity of details, such as a status update request which requires only the id of the job and the credentials of the initiator. Other jobs require more details, such as the data store scaling which requires a list of machines available to the initiator that can host data store instances, the current access point of the data store and credentials for the job initiator, for the data store and for the machines.

### 3.3.3 Security and Anonymization Engine

**Category Name**: security_engine
**Description**: An instance of this category will be created in order to have a security and privacy administrator that is global and always updated throughout the whole framework.
**Members**: The following members are defined for instances of the application control category.
- initiator_log: the initiator credentials for any access or modification in the security and privacy policies for a certain period of time, in JSON format.

**Links**: Security and Anonymization Engine instances will be linked to the Core component.
**Methods**: The standard OCCI methods are managed for this category endpoint with the following specificities.

POST

This method will be used to initiate a security and privacy job. This job can either be a modification to the current policies, an access control check or a security and privacy protocol query, in order for the Core to get information about the current protocols and enforce them.

PUT

The type of requests that Security and Anonymization Engine handles creates no need for continuous data transactions, thus the PUT method is not needed for this interface and it is disabled.

DELETE

This method may be used to delete a request and prohibit further access to all connected data.

### 3.3.4 Off-Loading APIs

**Category Name**: off_loading_apis
**Description**: Instances of this category will be created in order to communicate with and manage the individual data stores in the BASMATI federation.
**Links**: Off-loading APIs instances will be linked to the Core and Wrapper Modules.
**Methods**: The standard OCCI methods are managed for this category endpoint with the following specificities.

POST

This method will be used to initiate an off_loading_apis instance by providing the job type and the job details. These information will be forwarded from the Core component and as such we do not need the initiator credentials because basic authentication has already been concluded successfully at this point. What we need is the job ID in order to correlate the job response with the job request. This ID is added to the job_details JSON by the Core component.

### 3.3.5 Analytics Engine

**Category Name**: analytics_engine
**Description**: An instance of this category will be created in order to facilitate the communication between the BASMATI cloud and the individual Analytic components.
**Links**: Analytics Engine instances will be linked to the Analytic Modules.
**Methods**: The standard OCCI methods are managed for this category endpoint with the following specificities.

POST

This method will be used to initiate a data request by providing the job type (save or retrieve), the description of the data and the initiator credentials. As discussed, this category is functioning just as an interface to analytics modules in order to ensure access control and polyglot persistence to the monitoring data.

DELETE

This method may be used to delete a request and prohibit further access to all connected data.

## 4. USE CASES

## 4.1 Das Fest

Das Fest is a large three day event, taking place annually in Karlsruhe, Germany. A mobile application, called Das Fest App, is developed by YellowMap, providing a number of functions that support the operation of this event, enhancing the experience of the visitors and providing assistance when needed.

In detail, the application focuses on providing static and dynamic information to the visitors. The static information includes the event schedule, a map of the event area, information about the artists and the shows and other relevant information. The dynamic information includes features about locating friends, finding the nearest emergency spot or the toilet with the smallest queue and other information.

In this case, BUDaMaF is needed due to the high data demand. As this event is running since 1985, it attracts 200 to 400 thousand visitors each summer. The fact that the application is a mobile application, for IOS and Android, creates the need for a strong and responsive backend that is scalable in order to handle the few days of increased activity without bottlenecks, while saving resources the rest of the year. Even if a general idea of the visitors' number is estimated by previous years, the organizers cannot be sure of the attendance each year, thus they cannot be certain of how many resources they need.

BUDaMaF does not need apriori knowledge of the resource demand. It can contact the multi-cloud federation and find the most cost-effective solution for data store scaling and request offloading in semi real-time, countering bottlenecks as they arise, in the case of an underestimation of attendance. On the other hand, it can release resources to save on costs in the case of an overestimation of attendance, again in semi real-time. It also can be configured to favor data availability over disk space in its cost analysis process, performing finer cost balancing by switching cloud providers during the fluctuations of attendance.

## 4.2 Trip Builder

Trip Builder (Brilhante et al., 2014) is an unsupervised application, developed by CNR, that aims to assist tourists plan their visit in a city, taking into account budget and time limitations. On top of that, the application can automatically get information about the attractions available in any city and the opinion of their visitors about them, by crawling open internet sources like flickr, Wikipedia and twitter, thus it does not require manual registration of every attraction in a city.

In more detail, the application locates tourist hotspots from social media and other internet public sources and stores them as landmarks or attractions. Then a user can choose a city, her budget and the period of the visit in order to get an optimal path according to her choices and the opinion of the other users about the available landmarks, in order to visit the most popular attractions of the city in the most efficient way. This helps tourists unfamiliar with the city to make the most of their visit, even if they have a limited budget or a small time window.

TripBuilder is handling a lot of data, both crawled by external sources as already discussed and from its users that are constantly updating their preferences and their personal choices. All these data create a data management problem in the big data domain. The application is already based on a cloud architecture but in order for it to perform as expected, a chunk of the data need to follow the user. Given that the user is usually accessing the services from a mobile device, these data cannot be stored in the user's device. So, new data centers, near the user, need to be located and the data need to be moved to them while the user is travelling to the targeted city. This problem is similar to the one presented in the Virtual Desktop use case that we will discuss about in a following section.

## 4.3 Virtual Desktop

Mobile Virtual Desktop (MVD) (Kim et al., 2016) is an application provided by ETRI, a Korea based corporation. The main idea behind this application is having 24/7 access to a virtual computer system from any device that has internet access. This turns any device into a terminal, connected to a machine that can cover the needs of any user, be it a casual user that just browses the web or a power user that needs immense resources for heavy duty projects.

This functionality is achieved by binding resources in a cloud infrastructure in order to create a virtual machine using specifications collected by the user. Then the user can access the virtual machine through an user interface created by the application provider, that is installed on the user's device, be it a laptop, a smartphone, a tablet or any other device that has internet access and is supported by the interface.

The need for BUDaMaF arises in a couple of occasions but lets just present the most common one; when the user tries to move great distances. In this case, the user can encounter huge amounts of delays, every time a big data package needs to travel from the cloud infrastructure to the user interface or vice versa. As this application is made for all kind of devices, even smartphones, the data that can be kept in local memory or even stored in local hard drives is very limited. Thus, great amounts of data are always traveling back and forth, because the main bulk of the data are stored in the cloud.

When these big packages of data are travelling great distances, for example from Korea to Europe or vice versa, a delay is created, making the user interface unresponsive. This can be avoided if the data follow the user, starting a mitigation process while she is flying from Korea to Europe or vice versa. But in order for the application provider to support this operation, a cloud provider hosted in Europe has to be contacted in advance and an SLA signed. This cannot be done in real time, it needs days of preparation and market research and cost analysis. BUDaMaF is instead using the multi-cloud federation in order to find a low cost option in the area that the user will be and start transferring the data to a federation member the instant it gets notified that the user is flying to Europe.

## 5. FUTURE WORK

BUDaMaF is an ongoing project and as such there are still a lot to be done. The two main axis of future work in BUDaMaF concern the ACID enforcement and the creation of an intelligent agent. About the ACID enforcement we discussed in a previous chapter, mentioning that limited ACID support is already provided but efforts to provide full ACID support will be made.

The Intelligent Agent component, called BUDaMaFIA, will provide artificial intelligence to the framework. It will implement a machine learning model that locates and tries to predict load fluctuations and disastrous events and then tries to avoid or rectify them by using the BUDaMaF functionality. This agent will in fact replace the data administrator, by performing load balancing, data off-loading, security and privacy guideline management and other tasks automatically even before the need for such actions arise.

Of course, at the same time, the work on additional Wrappers and Analytics modules will continue, providing always improving data store support and analytics functionality.

## CONCLUSIONS

To conclude, we can see that BUDaMaF, even though a lot of work is still under way, is already a multi-function framework that can perform a plethora of tasks in any multi-cloud federation. Regardless of the underlying technologies and context of the federation, it can provide basic and advanced data management support, creating a polyglot persistent environment, as long as the federation resource manager provides a RestFul API for resource allocation. It can also provide a dashboard, allowing real-time, or near real-time, management of data and data stores in an environment of great complexity, such as a multi-cloud federation. Complexity that arises from the fact that cloud providers never aimed of working with one another or sharing their resources in a common pool.

## ACKNOWLEDGEMENTS

This work has been supported by the BASMATI project (http://www.basmati.cloud) and has been funded by the ICT R&D program of the Korean MSIP/IITP (R0115-16-0001) and the European Unions Horizon 2020 research and innovation programme under grant agreement no. 723131.